\documentclass[twocolumn,showpacs,amssymb,aps]{revtex4}
\usepackage{amsmath}
\usepackage{amsfonts}
\usepackage{amssymb}
\usepackage{graphicx}
\usepackage{epsfig}
\usepackage{dcolumn}
\usepackage{bm}

\def\lessim{\lower.5ex\hbox{$\; \buildrel < \over \sim \;$}}
\begin{document} \hbadness=10000
\topmargin -0.8cm
\preprint{}

\title{Enhanced Production of $\Delta$ and $\Sigma(1385)$  Resonances}
\author{Inga Kuznetsova and Johann Rafelski}
\affiliation{Department of Physics, University of Arizona, Tucson, Arizona, 85721, USA}
\affiliation{Department f\"ur Physik der Ludwig-Maximilians-Universit\"at M\"unchen und
Maier-Leibnitz-Laboratorium, Am Coulombwall 1, 85748 Garching, Germany}

\date{April 21, 2008}
%\maketitle
\begin{abstract}
Yields of    $\Delta(1230)$, $\Sigma(1385)$  resonances produced in
heavy ion collisions are studied within   the
framework of a kinetic master equation.   The  time evolution
is driven by the  process $\Delta \leftrightarrow N \pi $, $\Sigma(1385) \leftrightarrow \Lambda \pi $.  We obtain
  resonance yield  both below and  above chemical equilibrium, depending on initial
hadronization condition and separation of kinetic and chemical freeze-out.
\end{abstract}

%24.10.Pa Thermal and statistical models
%25.75.-q Relativistic heavy-ion collisions
%25.75.Nq Quark deconfinement,
%         quark--gluon plasma production and phase transitions in relativistic
%         heavy ion collisions
%12.38.Mh quark--gluon plasma  in quantum chromodynamics
 \maketitle

%\section{Introduction}
Hadron resonances are produced copiously in the quark-gluon plasma  (QGP)
fireball break up into hadrons (hadronization, chemical freeze-out)
e.g. at RHIC~\cite{Adams:2006yu,Salur:2006jq,Markert:2007qg,Witt:2007xa}.
Within the  statistical hadronization
model (SHM) approach~\cite{Torrieri:2004zz,Torrieri:2006xi},
the initial  yields are described by  chemical fugacities $\Upsilon$, and
hadronization temperature$\,T$. The production of heavy resonances
is suppressed exponentially in $m/T$.  Once  formed, resonances
decay. If this occurs inside matter, detailed balance requires also production
of resonances, called `regeneration' and/or `back-reaction'.

If the chemical freeze-out  occurs much earlier than thermal, the initially produced
resonances are practically invisible due to rescattering
of decay products~\cite{Rafelski:2001hp}. The observed yield of resonances
is fixed by the physical conditions prevailing at the  final  breakup of the fireball,
 at which time last scattering occurs, this is the  `kinetic freeze-out'.
The present work addresses two  questions:\\
a)  how observable  resonance
yield depends on the difference between chemical freeze-out temperature (e.g. point of hadronization of QGP)
 and  the kinetic freeze-out temperature;\\
b) how this yield depends on the degree of initial chemical non-equilibrium at hadronization. \\
One can see this work as an effort to improve on the concept of chemical freeze-out for
the case of resonances: given the relatively fast  reactions   their yield remains sensitive to  the
conditions prevailing  between chemical and thermal freeze-out, even if this time is just 1\,fm/c.

In order to describe evolution of the resonance abundance
one can  perform a microscopic transport simulation of the expanding system.
In this approach the  regeneration of resonances
was previously studied by Bleicher and collaborators~\cite{Bleicher:2002dm,Bleicher:2003ij,Vogel:2006rm}.
There are many detailed features of particle interactions to resolve in a microscopic model description
and thus it seems appropriate to   simplify the situation. We study  resonance decay and regeneration  
using the momentum integrated  population master equations, and assuming  hydrodynamic expansion
inspired model of fireball dynamics with conserved  entropy content.
In all our considerations we presume that
the yield of pions $\pi$    is so large
that we can assume it to be unaffected by any of the reactions we
consider, thus we fix pion yield in terms of fugacity and temperature values.
We do not consider  all resonances which decay into resonances
as does the SHARE2 SHM  program~\cite{Torrieri:2006xi},  thus we
will correct the final yields by an estimate of this effect comparing our initial resonance yield
to SHARE2 results.

For the `fast' baryon resonances considered here  we keep  the sum  yields  constant:
\begin{eqnarray}\label{partcon}
\Delta+ N&=&\Delta_0+N_0\equiv N^{\rm tot}_0 ={\rm Const.}, \\
\Sigma (1385)+  \Lambda & = &\Sigma_0(1385) + \Lambda_0\equiv \Lambda^{\rm tot}_0  ={\rm Const.} . \nonumber
\end{eqnarray}
The  baryon  annihilation,  strangeness exchange such as $N+K\leftrightarrow \Lambda+\pi$  reactions,
and population exchanges with   higher resonances
are assumed not to have a material impact within
the time scale during which the temperature drops from chemical to kinetic freeze-out condition.

The experimentally observable hyperon yield  appearing in our final result  is
\begin{equation} \label{Ltot}
\Lambda_{\rm tot}=\Sigma (1385)+ \Lambda+\Sigma^0(1193)  + Y^*
\end{equation}
due to experimentally inseparable  $\Sigma^0(1193) \to \gamma+\Lambda$ decay and
the decay of further hyperon resonances $Y^*$. Similarly, when
we refer to  $N_{\rm tot}$ we include   baryon resonances in the count.

%\section{$\Delta$ multiplicity evolution equation}

%\subsection{Equilibrium conditions for $\Delta$ density}

In the following   we will be referring explicitly to  the
$\Delta$  yield   governed by $ c\tau_\Delta\equiv 1/\Gamma_\Delta=1.67$ fm.
All equations apply equally to   $\Sigma(1385)  $ yield
(partial decay width $\Gamma_{\Sigma\to \Lambda}\simeq 35$ MeV) and we  will
compare our results   with experiment for this case.
We note that even though the $\Sigma(1385)  $ decay width is much smaller than $\Gamma_\Delta$, the
number of reaction channels and particle densities available lead to a significant
effect  for $\Sigma(1385)$, comparable to our finding for  $\Delta$.

The  evolution in time  of the $\Delta$ (or $\Sigma (1385)$) resonance yield is described
 by the process of resonance formation in scattering,
less natural decay:
\begin{equation} \label{delev}
\frac{1}{V}\frac{dN_{\Delta}}{dt}=\frac{dW_{N\pi \rightarrow
\Delta}}{dVdt}-\frac{dW_{\Delta \rightarrow N \pi}}{dVdt},
\end{equation}
where $N_{\Delta}$ is multiplicity of $\Delta$ resonances, ${dW_{N\pi \rightarrow \Delta}}/{dVdt}$ and ${dW_{\Delta
\rightarrow N \pi}}/{dVdt}$ are invariant rates (per unit volume and time) for $\Delta$
production and decay respectively.
Allowing for Fermi-blocking and Bose enhancement in the final state, the two in-matter rates are:
\begin{eqnarray}
\frac{dW_{\Delta \rightarrow N \pi }}{dVdt}=
&&\frac{g_{\Delta}}{(2\pi)^3}
     \int\frac{d^{3}p_{\Delta}}{2E_{\Delta}}f_{\Delta }
     \int\frac{d^{3}p_{N}}{2E_N(2\pi)^3}\left(1 - f_{N}\right)\nonumber\\[0.3cm]
&&\hspace*{-1.4cm}
\times\int\frac{d^{3}p_{\pi}}{2E_{\pi}\left(2\pi\right)^{3}}\left(1 + f_{\pi}\right)
\left(2\pi\right)^{4}\delta^{4}\left(p_{N}+p_{\pi}-p_{\Delta}\right)\nonumber\\[0.3cm]
&&\hspace*{-.4cm}
\times\frac{1}{g_{\Delta}}\sum_{spin}\left|\langle p_{\Delta}\left|
M\right|p_{N}p_{\pi}\rangle\right|^{2}\label{pd},
\end{eqnarray}
%and in analogy, we have for the $\Delta$ back-production rate
\begin{eqnarray}
\frac{dW_{\pi N \rightarrow \Delta}}{dVdt}&=&\frac{g_N}{(2\pi)^3}
\int\frac{d^{3}p_{N}}{2E_{N}}f_{N} \frac{g_{\pi}}{(2\pi)^3}
\int\frac{d^{3}p_{\pi}}{2E_{\pi}}f_{\pi}  \nonumber\\
&&\hspace*{-1.4cm}
\times \int\frac{d^{3}p_{\Delta}}{2E_{\Delta}\left(2\pi\right)^{3}}\left(1 - f_{\Delta}\right)
     \left(2\pi\right)^{4}\delta^{4}\left(p_{N}+p_{\pi}-p_{\Delta}\right)\nonumber\\
&&\hspace*{-.4cm}
\times\frac{1}{g_Ng_{\pi}}\sum_{\rm{spin}}\left|
\langle p_{N}p_{\pi}\left| M\right|p_{\Delta}\rangle\right|^{2}.
\label{dp}
\end{eqnarray}
where $g_i, i=\pi,N,\Delta$ is particles degeneracy. The distribution
functions for $\pi$, $N$,  $\Delta$ are
\begin{eqnarray}
f_{\pi}&=&\frac{1}{\Upsilon_{\pi}^{-1}e^{u\cdot p_{\pi}/T} - 1},\label{fpi}\\
f_{j}&=&\frac{1}{\Upsilon_j^{-1}e^{u\cdot p_j/T} + 1},\ j=N,\Delta\label{fN}.
\end{eqnarray}
Here $\Upsilon_i$ is particles fugacity, and  $u\cdot p_i=E_i$, for
$u^\mu=(1,\vec 0)$ in the rest frame of the heat bath where
 $d^4p\delta_0(p_i^2-m_i^2)\to d^3p_{i}/E_{i}$ for each particle. Hence,
Eq.(\ref{pd}) and Eq.(\ref{dp}) are  Lorentz invariant, as presented these rates
can be evaluated  in any convenient   frame  of reference. Normally, this is the
 frame co-moving with  the thermal volume element.

Eqs.(\ref{fpi}), (\ref{fN}) satisfy   the usual relations for the Fermi and Bose distributions:
\begin{eqnarray}
1 - f_{j}&=&\Upsilon_{j}^{-1} e^{u \cdot p_j}f_{j},\quad j=\Delta,N\\
1 + f_{\pi}&=&\Upsilon_{\pi}^{-1} e^{u \cdot p_{\pi}}f_{\pi}.
\end{eqnarray}
Using these equations,  we
obtain for decay rate Eq.(\ref{pd})
\begin{eqnarray}
&&\frac{dW_{\Delta \rightarrow N \pi
}}{dVdt}=\Upsilon_{N}^{-1}\Upsilon_{\pi}^{-1}\frac{1}{(2\pi)^6}
\int\frac{d^{3}p_{\Delta}}{2E_{\Delta}}\int\frac{d^{3}p_{N}}{2E_N} \nonumber\\
&&
\times\int\frac{d^{3}p_{\pi}}{2E_{\pi}\left(2\pi\right)^{3}}\left(2\pi\right)^{4}
\delta^{4}\left(p_{N}+p_{\pi}-p_{\Delta}\right) \nonumber\\
&&\times \sum_{spin}\left|\langle p_{\Delta}\left|
M\right|p_{N}p_{\pi}\rangle\right|^{2}f_{\Delta}f_{N}f_{\pi}\exp(u \cdot p_{\Delta}/T).\label{drg}
\end{eqnarray}
where in the last exponent we replaced $p_{N}+p_{\pi}$ by $p_{\Delta}$ given the energy-momentum
conservation 4-delta function. We can perform a similar simplification in Eq.(\ref{dp}). Then, observing
that due to  the time reversal symmetry,
$$\left|\langle p_{1}\left| M\right|p_{2}p_{3}\rangle\right|^2=\left|\langle p_{2}p_3\left| M\right|p_{1}\rangle\right|^2$$
we find a detailed balance relation between the production and decay rate:
\begin{equation}
{\Upsilon_{\Delta}}\frac{dW_{N\pi \rightarrow \Delta}}{dVdt}={\Upsilon_N \Upsilon_{\pi}}
          \frac{dW_{\Delta \rightarrow N \pi}}{dVdt}.\label{pdr}
\end{equation}

The master equation,  Eq.(\ref{delev}), can now be cast into the form:
\begin{equation}
\frac{1}{V}\frac{dN_{\Delta}}{dt}=
\left(\frac{\Upsilon_{\pi}\Upsilon_{N}}{\Upsilon_{\Delta}}-1\right)
    \frac{dW_{\Delta\rightarrow N \pi}}{dVdt}.\label{ddup}
\end{equation}
This is a rather  intuitive and simple result, yet only recently the $1\leftrightarrow 2$ population master equations
have been considered~\cite{KuznKodRafl:2008}.
Equation (\ref{ddup}) implies  for $dN_\Delta/dt=0$ the   chemical equilibrium condition:
\begin{equation}
\Upsilon^{\rm eq}_{\pi} \Upsilon^{\rm eq}_{N}=\Upsilon^{\rm eq}_{\Delta}. \label{equilcon}
\end{equation}
This equation is solved by the global chemical equilibrium
$\Upsilon^{\rm eq}_{\pi} =\Upsilon^{\rm eq}_{N}=\Upsilon^{\rm eq}_{\Delta}=1$.
However, there are also other, transient, equilibrium states possible,
given a  prescribed value of e.g. the background
pion abundance, $\Upsilon^{\rm eq}_{\pi} \ne 1$.  When the initial state is formed
away from transient equilibrium condition, we recognize that for
$\Upsilon_{\Delta}<\Upsilon_{\pi} \Upsilon_{N}$  the
$\Delta$ production is dominant, and conversely,
for $\Upsilon_{\Delta}>\Upsilon_{\pi} \Upsilon_{N}$  the $\Delta$ decay
dominates.

We now introduce into the population master equation (\ref{ddup})
the effective lifespan,  $\tau_\Delta$ aiming to find an equation
similar to classic radioactive decay population equation. We define
the in medium $\Delta$-lifespan to be:
\begin{equation}\label{Delt}
\tau_\Delta\equiv \frac{\Upsilon_\Delta}{V} \frac{dN_\Delta/d\Upsilon_\Delta}{dW_{\Delta\to N\pi}/dVdt}.
\end{equation}
We recognize that in the Boltzmann limit this corresponds to the ratio of equilibrium yield to the rate per unit time
at which  the equilibrium is approached.
We obtain  for  Eq.(\ref{ddup}):
\begin{equation}\label{Ndelta}
\frac{dN_{\Delta}}{dt}=\left (\Upsilon_{\pi}\Upsilon_{N}- \Upsilon_{\Delta} \right)
              \frac{dN_{\Delta}}{d\Upsilon}\frac 1 {\tau_\Delta}.
\end{equation}

In case that
the ambient temperature does not vary with time, and thus only populations evolve due to change in
fugacities, we have  $dN/dt=dN/d\Upsilon\, d\Upsilon/dt$ and
 the following dynamical equation for the fugacity arises:
\begin{equation}\label{dUpsdt}
\tau_\Delta\frac{d\Upsilon_\Delta}{dt}=\left (\Upsilon_{\pi}\Upsilon_{N}- \Upsilon_{\Delta} \right).
\end{equation}
This is `classical' population equation form where the fugacity plays the role of the classical densities.
When the dynamical values of  $\Upsilon_i(t)$ are used in the quantum Bose/Fermi distributions, the
effects of blocking, and stimulated emission are explicit.

If we instead  were to introduce the lifespan by  $\tilde
\tau_\Delta\equiv (N_\Delta/V)/(dW_{\Delta\to N\pi}/dVdt)$,  this
implies for all particles (Bose, Fermi, Boltzmann)  the classical
population equation, e.g. $dN_\Delta/dt
=(\Upsilon_{\pi}\Upsilon_{N}/\Upsilon_{\Delta}
-1)N_\Delta/\tilde\tau_\Delta$, and the quantum effects are now
hidden in the  definition of $\tilde \tau$. Both definitions
coincide for the case of a dilute system, and differ most for dense
systems. In the limit of very dilute, vacuum system, the relaxation
time is the same as the lifespan of the particles. The computed
yields of particles as function of time are  not dependent on the
finesse of the relaxation time definition.

We now set up for semi-analytical solution of master equation (\ref{Ndelta}).
For multiplicities ${\Delta}$ and ${N}$ considering the small yield and $m\gg T$
we will use the  Boltzmann distribution:
\begin{eqnarray}
\frac {N_{\Delta} }V&=&\Upsilon_{\Delta}\frac{T^3}{2\pi^2}g_{\Delta}x_{\Delta}^2K_2(x_{\Delta}), \\
\frac {N_{N} }V&=&\Upsilon_{N}\frac{T^3}{2\pi^2}g_{N}x_{N}^2K_2(x_{N}),
\end{eqnarray}
where $x_{\Delta,N }=m_{\Delta,N }/T$, $K_2(x)$ is Bessel function.
Considering that  fugacities, temperature  and volume vary in
time, we   rewrite the left hand side of Eq.(\ref{Ndelta}):
\begin{equation}
\frac{\ dN_{\Delta}}{N_\Delta d\tau} = \frac{\ d\Upsilon_{\Delta}}{\Upsilon_{\Delta}d\tau } +
 \frac{d\ln(x_{\Delta}^2K_2(x_{\Delta}))}{dT}\dot{T} + \frac{d(VT^3)}{VT^3\ \ d\tau}.
\label{Ups}
\end{equation}
We changed from $t$ to $\tau$ to make explicit the fact that we work in fluid-element co-moving frame
and thus do not consider the effect of flow on the volume time dependence.

Combining  Eq.(\ref{Ndelta}) with Eq.(\ref{Ups}) we obtain
\begin{equation}
\frac{d\Upsilon_{\Delta}}{d{\tau}} =
\left({\Upsilon_{\pi}\Upsilon_{N}}-\Upsilon_{\Delta}\right)\frac{1}{\tau_{\Delta}}
+\Upsilon_{\Delta}\frac{1}{\tau_T}+\Upsilon_{\Delta}\frac{1}{\tau_S} ,
\label{Ups2}
\end{equation}
%where
\begin{eqnarray}
\frac{1}{\tau_{T} } &=&  -\frac{d\ln({x_{\Delta}}^2K_2(x_{\Delta}))}{dT} \dot{T}. \label{Teq}\\
\frac{1}{\tau_{S} } &=&  -\frac{d\ln( VT^3  )}{dT} \dot{T}. \label{Seq}
\end{eqnarray}
The last term is negligible, $\tau_S\gg \tau_\Delta, \tau_T$ since pions dominate and we have near conservation of
entropy which for massless particles would in fact imply $VT^3=$Const.

Since entropy must be (slightly) increasing, while $T$ is decreasing with time, $\tau_S>0$. Similarly,
$\tau_T>0$, since the temperature decreases with time, and $x^2K_2(x), x=m/T$ increases with $T$ :
\begin{equation}
x^2K_2(x) \approx \sqrt{0.5 \pi}x^{3/2}\exp(-x);
\end{equation}
Therefore:
\begin{equation}\label{1tT}
\frac{1}{\tau_T} \approx -\frac{m_{\Delta}}{T}\left( 1-\frac 3 2 \frac T m_{\Delta}\ldots \right)\frac{\dot{T}}{T}.
\end{equation}

We now evaluate the magnitude of $\tau_T$   invoking a model of matter expansion
of the type used e.g. in~\cite{Letessier:2006wn}, where the longitudinal
and transverse expansion is considered to be independent.
In the proper rest frame of the outflowing matter,
\begin{equation}
\frac{dS}{dy}\propto T^3\frac{dV}{dy}= \pi R^2_{\perp}(\tau)T^3\frac {dz}{ dy}\simeq {\rm Const.}. \label{volt1}
\end{equation}
We will use $  {dz}/{ dy}\simeq \tau$, where $\tau$ is the proper time in the
local volume element,  this  is exact for a 1-d ideal hydro flow.
The growth of the transverse dimension
can be generically described by:
\begin{equation}
R_{\perp}(\tau) = R_0 + \int^\tau_{\tau_0} v(\tau')d{\tau'}.
\end{equation}
From  Eq.(\ref{volt1}) by elementary evaluation we obtain:
\begin{equation}\label{DTT}
\frac{\dot {T}}{T} = -\frac{1}{3}\left( \frac{2\,(v\tau/R_{\perp  })  + 1}{  \tau}\right).
\end{equation}

\begin{figure}
\centering
\includegraphics[width=8.3cm]{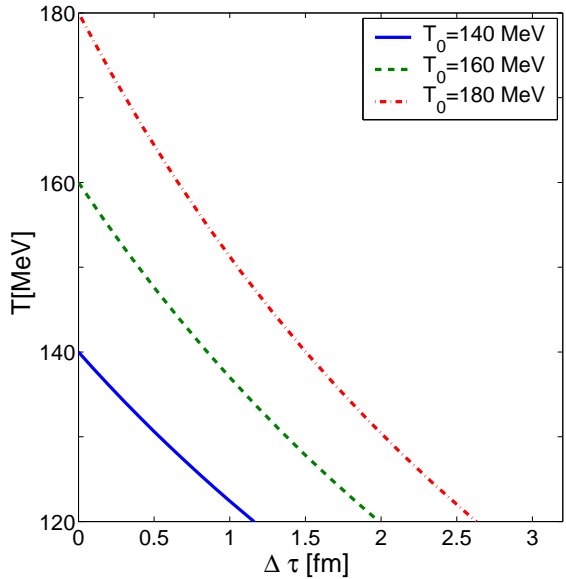}
\caption{\small{Temperature $T$ as function of $\delta \tau$, the proper time interval between
chemical and thermal freeze-out or  chemical
freeze-out temperature (from top to bottom)
 $T=180,\, 160,\, 140$ MeV and thermal  freeze-out $T\ge 120$ MeV.} }
\label{Ttaut}
\end{figure}

Equation (\ref{DTT}) evaluated near hadronization condition is yielding the magnitude of $\tau_T$, see Eq.(\ref{1tT}).
If the maximum expansion velocity is practically instantly achieved, $v\tau/R_\perp\simeq 1$. This leads to maximum
value of $\dot T /T\simeq -1/\tau$. However if a more realistic profiles are assumed, $\dot T /T$ is diminished
in magnitude as much as 30\%. We thus conclude that
$$  \frac {0.5}{\tau_h}\, \frac{m_{\Delta}}{T}< \frac 1 {\tau_T} < \frac {0.7}{\tau_h}  \frac{m_{\Delta}}{T}$$
which for hadronization time $\tau_h<10$ fm can compete with  the width of the $\Delta$-resonance, $1/t_\Delta\simeq 120$ MeV.
As this shows, the details of the expansion model are not critical for the results we obtain. In actual calculations we employ
$v(\tau)$ described in~\cite{Letessier:2006wn}, where we assume that the expansion is already at maximum velocity
at the time of chemical freeze-out. The resulting dependence $T(\tau)$ after chemical freeze-out is shown in figure \ref{Ttaut}.
We note that the time between chemical and thermal freeze-out $\Delta \tau$ is not longer than about 2.5fm/c, and can be as short
as  1fm/c. However, even such a short scattering period is enough to alter the visible yields of strong resonances, in fact most 
pronounced effect we find in the latter case, since the longer time allows a greater degree of chemical equilibration.

%%%%%%%%%%%%%%%%%%%%%%%%%%%%%%
%%%%%%%%%%%%%%%%%%%%%%%%%%%

We now can solve  Eq.(\ref{Ups2}). Employing Eq.(\ref{partcon}) we have:
\begin{equation}
\frac{d\Upsilon_{\Delta}}{d\tau} + \tilde\Gamma (\tau)\Upsilon_{\Delta} = q(\tau), \label{Ups3}
\end{equation}
%where
\begin{eqnarray}
\tilde\Gamma(\tau) &=& \left[1+\Upsilon_{\pi}\frac{N^{\infty}_{\Delta}}{N^{\infty}_{N}}\right]\frac{1}{\tau_{\Delta}} - \frac{1}{\tau_T}, \\[0.3cm]
q(\tau) &=& \Upsilon_{\pi}\frac{N_0^{\rm tot}}{N^{\infty}_{N}}
\frac{1}{\tau_{\Delta}},
\end{eqnarray}
where $N^{\infty}_{\Delta}$ and $N^{\infty}_{N}$ are densities of
$\Delta$ and $N$ resonances with $\Upsilon_{\Delta}=\Upsilon_N=1$.
The solution of Eq.(\ref{Ups3}) is elementary:
\begin{eqnarray}
\Upsilon_{\Delta} (\tau)= \left(\Upsilon^0_{\Delta} +
     \int_{\tau_{h}}^{\tau}q\,e^{ \int_{\tau_{h}}^{\tau'}\tilde\Gamma d\tau''}d\tau' \right )e^{ -\int_{\tau_h}^{\tau }\tilde\Gamma  d\tau' }
\end{eqnarray}
where $\tau_h$ is initial expansion time at hadronization, and $\tau_h<\tau<\tau_{\rm max}$, upper time limit chosen
to yield  $T_ {\rm max}= 120$ MeV, i.e. $\tau_{\rm max}\simeq 8$ fm.

%%%%%%%%%%%%%%%%%%%%%%%%%%%%%%%%%%%%%%%%%%%%%%%%%%%%%%%%%%%%%%%%%%%%%%%%%%%%%%%%%%%%
%\section{Results for $\Delta$ resonance multiplicity}

In order to evaluate the  final $\Delta$ multiplicity we need also to know initial
particles densities  right after hadronization which we consider for RHIC
head-on Au--Au collisions at $\sqrt{s_{\rm NN}}=200$ GeV.
We introduce  the  initial hadron yields  inspired by a picture of a rapid
hadronization of QGP  with all hadrons produced with
yields governed by entropy and strangeness
content of QGP by quark recombination. In this model
the yields  of mesons and baryons are controlled by
the  constituent  quark fugacity $\gamma_q$:
\begin{equation}
\Upsilon^0_{\pi}=\gamma_q^{2}; \qquad \label{upinpi}
\Upsilon^0_{\Delta,N}=\gamma_q^{3}. %\label{upindN}
\end{equation}
Thus for $\gamma_q>1$ we have the condition $\Upsilon_{\Delta}<\Upsilon_{\pi} \Upsilon_{N}$,
and the yield of $\Delta$ will increase  in the time evolution.

For each
entropy content of the QGP fireball, the corresponding fixed background value of $\gamma_q$ can
be found once hadronization temperature is known~\cite{Kuznetsova:2006bh}.  For $T=140$ MeV pions
form a  nearly fully  degenerate Bose gas with $\gamma_q\simeq 1.6$.
In the following discussion, aside of this  initial condition,  we also consider the value pairs
$T=150{\rm\, MeV},\,\gamma_q=1.42$,
 $T=160{\rm\, MeV},\,\gamma_q=1.27$,
  $T=170{\rm\, MeV},\,\gamma_q=1.12$   and
$T=180$ MeV with $\gamma_q=1$.
Note further that
for  $m\gg T$ the density   $\Delta$ is relatively low,  thus there is no significant dependence of $1/{\tau_{\Delta}}$ on
$T$   and $\Upsilon_{\Delta}$; in essence $\tau_\Delta=\hbar /\Gamma_\Delta$
 takes the free space value $\tau_\Delta\simeq \hbar/120\rm{MeV}$.

%%%%%%%%%%%%%%%%%%%%%%%%%%%%%%
\begin{figure}
\centering
\includegraphics[width=8.3 cm,height=8.3 cm]{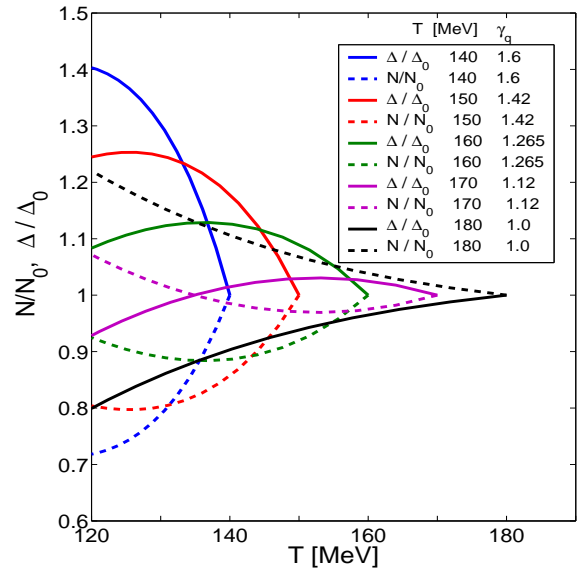}
\caption{\small{The ratio ${\Delta}/{\Delta}^0$ (solid lines) and $N/N^0$ (dashed lines)
 as functions of temperature $T$ for select given pairs of values $T,\gamma_q$,
see text and figure box for details.}}
\label{nd}
\end{figure}
%%%%%%%%%%%%%%%%%%%%%%%%%%%

In figure \ref{nd} we present results for ratios
${\Delta}/{\Delta}_0$ (solid lines) and $N/N_0$ (dashed lines) as
functions of temperature $T$, beginning from the presumed initial hadronization
temperature $T$ through  $T_ {\rm max}= 120$ MeV.
${\Delta}_0$ and $N_0$ are the initial yields obtained at each hadronization
temperature.  For $T < 180$ MeV, initially $\Upsilon_\Delta < \Upsilon_N\Upsilon_\pi$,
thus   based on our prior discussion, we expect that
the master equation leads to an initial increase in the yield of resonances. However,
as temperature drops,  due to the dynamics of the expansion   the increasing yield
of $\Delta$ turns over, and a final nett
increase of resonance yield is observed   for $T\le 160$ MeV. We note that for $T \ge 180$ MeV there is 
a  continuous depletion of resonance yield. The nucleon yields
move in opposite direction to the $\Delta$-resonance.

This behavior can be understood in qualitative manner as follows:
  The total number $\Delta + N$ is conserved therefore $\Delta$ multiplicity increases and $N$
multiplicity decreases until they reach transient chemical equilibrium
($dN_{\Delta}/d\tau=0$),  corresponding to the maximum point seen for  $\Delta$ in figure \ref{nd}.
There is also influence of expansion:  even if for some temperature
the transient equilibrium condition (\ref{equilcon}) is reached,  the system cannot
stay in this equilibrium, $\Upsilon_{\Delta}$ and $\Upsilon_{N}$ are increasing to conserve
total number of particles. $\Upsilon_{\Delta}$ increases faster
because of larger $\Delta$'s mass.  After
$\Upsilon_{\Delta}$ becomes larger than $\Upsilon_{\pi}\Upsilon_{N}$
$\Delta$ decay begins to dominate and  their multiplicity is decreasing. The special case at
hadronization temperature $T=180$ MeV where, $\Upsilon_i=1$ and equilibrium
condition is satisfied initially. As expansion sets in,   ${\Delta}$ is decreasing
because $\Upsilon_{\Delta} >  \Upsilon_{N}$ (recall that here $\Upsilon_{\pi}=1$).  
In the SHM evaluation of yields one assumes that all ratios seen in figure \ref{nd}  are unity.

\begin{figure}
\centering
\includegraphics[width=8.3 cm,height=8.3 cm]{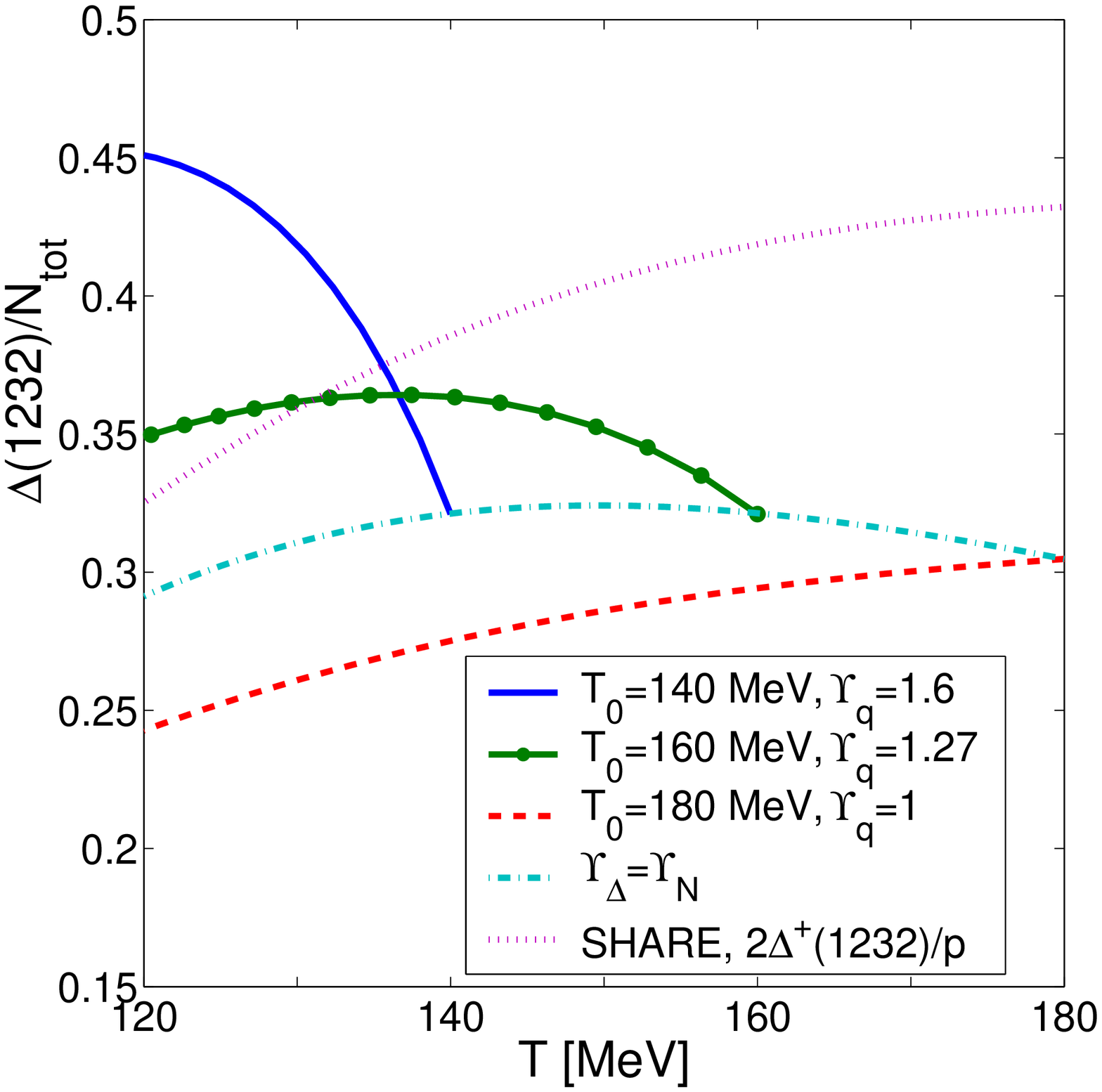}\\[-0.6cm]
\includegraphics[width=8.3 cm,height=8.3 cm]{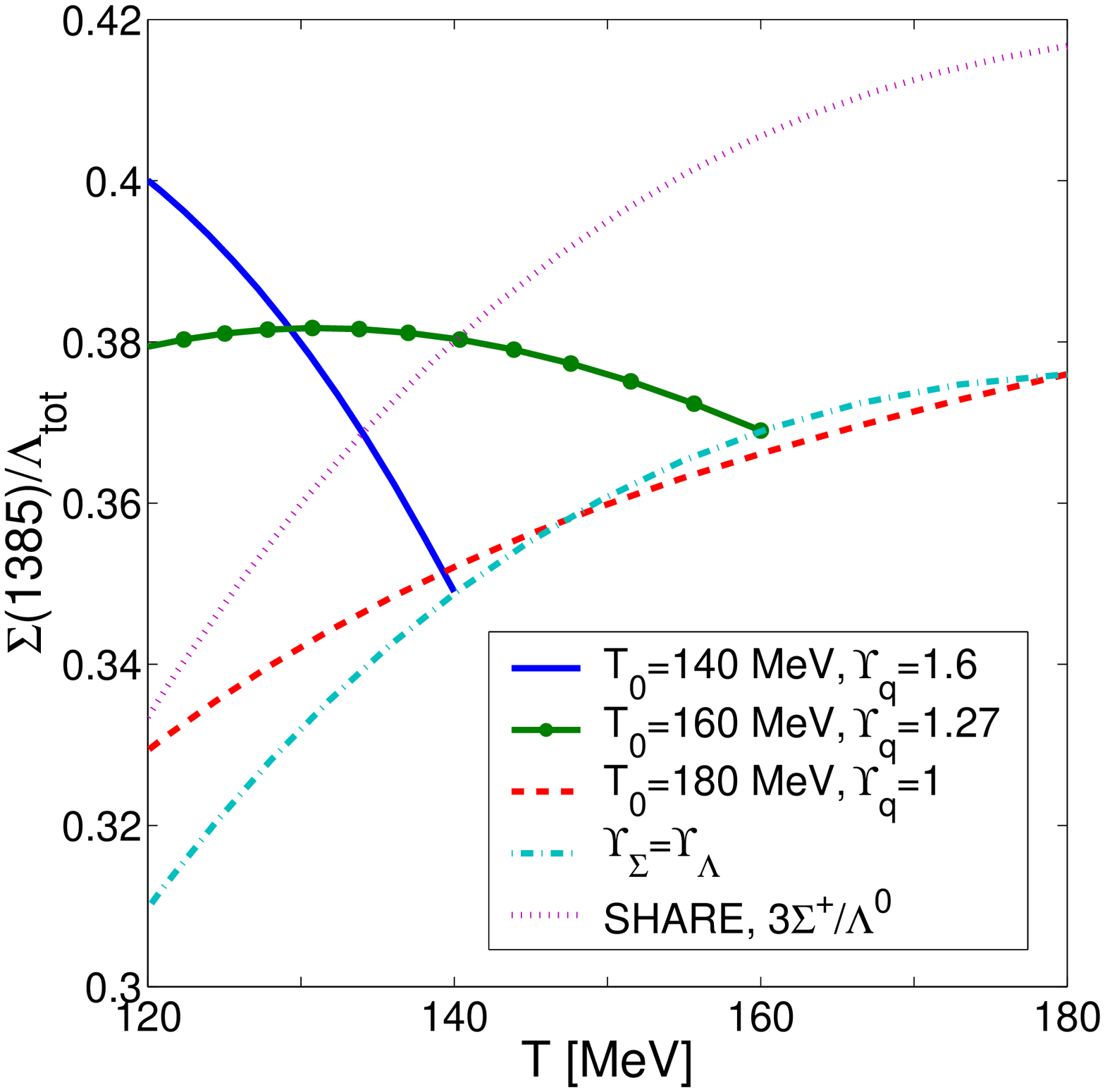}
\caption{\small{Relative resonance yield, for (top)
${\Delta}/N_{\rm tot}$ and (bottom)
${\Sigma(1385)/\Lambda_{\rm tot}}$ as a functions of freeze-out
temperature, for hadronization temperatures
$T_0=140,160,180$ MeV, see box and text for details.
The dotted brown line gives the expected SHM chemical equilibrium  result.} }
\label{ndNtot}
\end{figure}

The initial hadronization  yields which we used as
reference in   figure \ref{nd} are not accessible to measurement. Therefore,
we  consider in  figure \ref{ndNtot} the fractional yield   ${\Delta}/N_{\rm tot}$ (top frame), again as a
function of temperature $T$. The results  for hadronization temperatures $T_0=140$ (solid
blue line), $T_0=160$ (dash-dot green line) and $T_0=180$ MeV (dashed brown
line) are shown. $N_{\rm tot}$ is fixed by hadronization
condition and is not a function of time, as discussed.  Thus the   observable final rapidity nucleon yield
corresponds to the initial  value at hadronization. Note that 
up to strange and multi  strange baryon contribution,  $N_{\rm tot}$
is the  total baryon (rapidity) yield.

%%%%%%%%%%%%%%%%%%%%%%%%%%%%%%
%%%%%%%%%%%%%%%%%%%%%%%%%%%

Since in this study  we have considered a  subset of all relevant baryon resonances
our chemical equilibrium reference yield (line for $\Upsilon_{\Delta}=\Upsilon_{N}$) is not the same as
the corresponding reference line for the full statistical hadronization model (SHM)
evaluation,  obtained using SHARE2, and presented as    $2\Delta^+/p$ (upper frame) and
 $3\Sigma^+/\Lambda^0$ (lower frame).
 The SHARE2-SHM value  $\Delta^{++}/p\simeq 0.2$ at $T\simeq 160$ MeV
is consistent with the STAR d--Au results~\cite{Abelev:2008yz}. Also,  comparing
our with the SHARE2  result we note that SHARE2  yield
 is larger at chemical freeze-out. The magnitude of
the difference in the yields at time of chemical freeze-out provides a
measure of the magnitude of the corrections we can expect to arise in
the full treatment at thermal freeze-out and/or systematic error for these
yields.

The nature of these effects is  different for the two yield cases considered: the
presence of heavier  resonances which cascade by way of $\Delta$ leads to an increase
of the thermal freeze-out yield. The correction is thus nearly as much as we see the SHARE2 yield higher at
chemical freeze-out.
For $\Sigma(1385) $  the difference  with SHARE2 arises from a difference  of contributions
of partial decays producing $\Lambda_{\rm tot}$, thus the correction is multiplicative
factor which does not change, but is uncertain in magnitude due to lack of
knowledge about the branching ratios.

We believe that  the  $\Delta$ and  $\Sigma(1385) $ yields are  underestimated  by about 15\% -- 35\%.
(bigger effect for hadronization at higher $T$).
This implies that depending on hadronization temperature   a relative yield range
$0.16<\Delta^{++}/p=0.5 {\Delta}/N_{\rm tot}<0.26$ arises,
and similarly (see lower frame in figure \ref{ndNtot})   $0.35<\Sigma(1385)/\Lambda_{\rm tot}<0.43$
with the {\it higher} relative yield
corresponding to the {\it lower} hadronization temperature. One of the key results of this work
is the narrow range for $\Sigma(1385)/\Lambda_{\rm tot}$, and the
fact that the initial chemical non-equilibrium  effect leads to a
reversal of  the SHM model situation:   the relative yields of massive resonances decreases
with decreasing hadronization temperature.  .

In order to compare with the experimental results we
note that the data presented~\cite{Adams:2006yu,Salur:2006jq}
are for  charged $\Sigma(1385)$,
particle and antiparticle channels,
$(\Sigma^\pm(1385)+\overline{\Sigma^\pm(1385)}/(\Lambda_{\rm tot}+\overline{\Lambda_{\rm tot}})\simeq 0.29$.
This result needs to be multiplied with 3/2 to be comparable to results presented here which include $\Sigma^0(1385)$.
Multiplying the  value   for hadronization at $T=140$ MeV with thermal
freeze-out at $T=120$ MeV, and  allowing for contribution by heavier resonances as indicated by SHARE2
our result is in perfect agreement with~\cite{Adams:2006yu,Salur:2006jq}
However, given the narrow range of results we find, it seems that the high yield of $\Sigma(1385)$, seen
the error ${\cal O}(20\%)$ is nearly compatible with the entire range of chemical freeze-out
temperatures here considered -- the low $T$ chemical freeze-out is favored by 1.5  s.d. over high $T$.

The reader should take note that the `thermal' model result presented in Ref.~\cite{Adams:2006yu}
corresponds to initial high temperature freeze-out in chemical equilibrium   which is
 unobservable, since the high $T$ hadronization resonance decay products have no chance
to escape into free space. Thus this comparison of this model with experiment is flawed.
The evolved yield is shown as (red) dashed line in figure \ref{ndNtot}, and is found 25\%
below the value measured. The reason this happens is that the high $T$ chemical freeze-out
happens near chemical equilibrium and the yields follow closely the chemical equilibrium 
yield described by temperature, thus it is the
{\em thermal freeze out temperature which in this case controls the final observable
resonance yield}.

In summary,
we have presented master equation governing the evolution in time
of the $\Delta, \Sigma(1385)$ baryon resonance yield after QGP hadronization, allowing
for  resonance decay and production process.
We have shown  considering the properties of the master equation that if the
yield of hadrons is initially above chemical equilibrium,
the resonance population increases beyond the initial   yield.
Conversely, we find that in a physical system  in which the particle
multiplicities  of hadrons arise   below chemical equilibrium yields, a circumstance
expected below threshold to QGP formation, the final  yield of resonances is
suppressed by  the  dominance of the resonance decay process over back reaction  resonance production.

In a quantitative model we evolved the yields
after QGP hadronization allowing for initial chemical
non-equilibrium particle abundances,  and volume expansion assuring  entropy
conservation. We found, see   figure  \ref{ndNtot}, that the thermal freeze-out fractional resonance yield
differs significantly from the chemical-freeze out SHM expectation, with the scenario involving
high-$T$ hadronization resonance yield being depleted,
and low-$T$ hadronization yield  scenario further enhanced in relation to the total  yield.

The resonance enhancement effect  we presented
can only occur when the initial state is out of chemical equilibrium, and the decay/formation processes
are fast enough to compete
with  the hadron volume evolution. One would thus think that  `narrow', i.e. quasi-stable resonances
are not subject to the effects considered here. However, a special consideration
must be given to   narrow resonance which are  strongly coupled to more massive resonances
which can decay fast into other channels. An example is  $\Lambda(1520)$. In this situation
the lower mass state, here $\Lambda(1520)$,  is the ground state which is depleted
by coupling to the yet more massive state. There
are several states of relevance to consider, hence $\Lambda(1520)$ `quenching'~\cite{Rafelski:2001hp}
must be addressed in a more rigorous numerical approach.
We  will  to return to  discuss this case in a separate publication \cite{KRprep}.

Aside of several
specific prediction we made here,  there are three important  general consequences of our study: a)
the fractional yield of resonances $A^*/A$ can be considerably higher than expected naively
in SHM model of QGP hadronization,   b) since there is nearly
a factor two difference in the final thermal freeze-out ratio in $\Delta/N_{\rm tot}$, while the SHM yields
  a more $T$ independent result,  one can imagine the use of $\Delta/N_{\rm tot}$ as
a tool to distinguish the different hadronization conditions e.g. chemical non-equilibrium vs chemical equilibrium
a point noted in similar context before~\cite{Torrieri:2006yb}; and c) we have
shown that the relatively high yield of charged  $\Sigma^\pm(1385)$
reported by STAR is  well explained  by our considerations
with  hadronization at $T=140$ MeV being favored.

\subsubsection*{Acknowledgments}
This research was supported
by a grant from: the U.S. Department of Energy  DE-FG02-04ER4131;
and  by the DFG Cluster of Excellence MAP --
Munich Center for Advanced Photonics. 
These results were first presented at the 24th Winter
Workshop on Nuclear Dynamics on April 6, 2008, see:
http://rhic.physics.wayne.edu/\~{ }bellwied/wwnd08/ kouznetsova-wwnd08.ppt

%\vspace*{.2cm}
%\subsubsection*{Acknowledgments}
%{\it Work supported by a grant from: the U.S. Department of Energy  DE-FG02-04ER4131}

%%%%%%%%%%%%%%%%%%%%%%%%%%%%%%%%%%%%%%%%%

%%%%%%%%%%%%%%%%%%%%%%%%%%%%%%%%%%%%%%%%%
\vspace*{-0.3cm}
%%%%%%%%%%%%%%%%%%%%%%%%%%%%%%%%%%%%%%%%%
%\begin{references}


\begin{thebibliography}{19}
\providecommand{\bibinfo}[2]{#2}
%\vspace*{-0.3cm}
%\cite{Salur:2006jq}

%\cite{Adams:2006yu}
\bibitem{Adams:2006yu}
  J.~Adams {\it et al.}  [STAR Collaboration],
  %``Strange baryon resonance production in s(NN)**(1/2) = 200-GeV p + p and  Au
  %+ Au collisions,''
  Phys.\ Rev.\ Lett.\  {\bf 97}, 132301 (2006)
  [arXiv:nucl-ex/0604019].
  %%CITATION = PRLTA,97,132301;%%

%\cite{Salur:2006jq}
\bibitem{Salur:2006jq}
  S.~Salur,
  %``Baryonic resonance studies with STAR,''
  J.\ Phys.\ G {\bf 32}, S469 (2006)
  [arXiv:nucl-ex/0606002].
  %%CITATION = JPHGB,G32,S469;%%

%\cite{Markert:2007qg}
\bibitem{Markert:2007qg}
  C.~Markert  [STAR Collaboration],
  ``Resonance production in heavy-ion collisions at STAR,''
  arXiv:0712.1838 [nucl-ex]., J. Phys. G (in press) (2008).
  %%CITATION = ARXIV:0712.1838;%%

%\cite{Witt:2007xa}
\bibitem{Witt:2007xa}
  R.~Witt,
  %``Xi(1530)0 production in heavy-ion collisions and its implications for
  %Delta(t(therm-chem)),''
  J.\ Phys.\ G {\bf 34}, S921 (2007)
  [arXiv:nucl-ex/0701063].
  %%CITATION = JPHGB,G34,S921;%%


%\cite{Torrieri:2004zz}
\bibitem{Torrieri:2004zz}
  G.~Torrieri, S.~Steinke, W.~Broniowski, W.~Florkowski, J.~Letessier and J.~Rafelski,
  %``SHARE: Statistical hadronization with resonances,''
  Comput.\ Phys.\ Commun.\  {\bf 167}, 229 (2005)
  [arXiv:nucl-th/0404083].
  %%CITATION = NUCL-TH 0404083;%%

%\cite{Torrieri:2006xi}
\bibitem{Torrieri:2006xi}
  G.~Torrieri, S.~Jeon, J.~Letessier and J.~Rafelski,
  %``SHAREv2: Fluctuations and a comprehensive treatment of decay feed-down,''
  Comput.\ Phys.\ Commun.\  {\bf 175}, 635 (2006)
  [arXiv:nucl-th/0603026].
  %%CITATION = CPHCB,175,635;%%


%\cite{Rafelski:2001hp}
\bibitem{Rafelski:2001hp}
  J.~Rafelski, J.~Letessier and G.~Torrieri,
  %``Strange hadrons and their resonances: A diagnostic tool of QGP  freeze-out
  %dynamics,''
  Phys.\ Rev.\  C {\bf 64}, 054907 (2001)
  [Erratum-ibid.\  C {\bf 65}, 069902 (2002)]
  [arXiv:nucl-th/0104042];\\
  %%CITATION = PHRVA,C64,054907;%%
%\cite{Torrieri:2001ue}
%\bibitem{Torrieri:2001ue}
G.~Torrieri and J.~Rafelski,
  %``Strange hadron resonances as a signature of freeze-out dynamics,''
  Phys.\ Lett.\  B {\bf 509}, 239 (2001)
  [arXiv:hep-ph/0103149.
  %%CITATION = PHLTA,B509,239;%%

%\cite{Bleicher:2002dm}
\bibitem{Bleicher:2002dm}
  M.~Bleicher and J.~Aichelin,
  %``Strange resonance production: Probing chemical and thermal freeze-out  in
  %relativistic heavy ion collisions,''
  Phys.\ Lett.\  B {\bf 530} (2002) 81
  [arXiv:hep-ph/0201123].
  %%CITATION = PHLTA,B530,81;%%


%\cite{Bleicher:2003ij}
\bibitem{Bleicher:2003ij}
  M.~Bleicher and H.~Stoecker,
  %``Dynamics and freeze-out of hadron resonances at RHIC,''
  J.\ Phys.\ G {\bf 30}, S111 (2004)
  [arXiv:hep-ph/0312278].
  %%CITATION = JPHGB,G30,S111;%%

%\cite{Vogel:2006rm}
\bibitem{Vogel:2006rm}
  S.~Vogel and M.~Bleicher,
  %``Resonance production in heavy ion collisions - what can we learn from
  %RHIC?,''
  arXiv:hep-ph/0607242; in proceedings of ``22nd Winter Workshop on Nuclear Dynamics''
La Jolla, CA, 11-19 March, 2006.
  %%CITATION = HEP-PH/0607242;%%


%\cite{KuznKodRafl:2008}
\bibitem{KuznKodRafl:2008}
  I. Kuznetsova, T. Kodama and J.~Rafelski,
  ``Chemical Equilibration Involving Decaying Particles at Finite
  Temperature ''
  in preparation.



%\cite{Kuznetsova:2006bh}
\bibitem{Kuznetsova:2006bh}
  I.~Kuznetsova and J.~Rafelski,
  %``Heavy flavor hadrons in statistical hadronization of strangeness-rich
  %QGP,''
  Eur.\ Phys.\ J.\  C {\bf 51}, 113 (2007)
  [arXiv:hep-ph/0607203].
  %%CITATION = EPHJA,C51,113;%%

%\cite{Letessier:2006wn}
\bibitem{Letessier:2006wn}
  J.~Letessier and J.~Rafelski,
  %``Strangeness chemical equilibration in QGP at RHIC and LHC,''
  Phys.\ Rev.\  C {\bf 75}, 014905 (2007)
  [arXiv:nucl-th/0602047].
  %%CITATION = PHRVA,C75,014905;%%


%\cite{Abelev:2008yz}
\bibitem{Abelev:2008yz}
  B.~I.~Abelev {\it et al.}  [STAR Collaboration],
  ``Hadronic resonance production in $d$+Au collisions at $\sqrt{s_{_{NN}}}$ =
  200 GeV at RHIC,''
  arXiv:0801.0450 [nucl-ex]. %, submitted to Phys. Rev. C.
  %%CITATION = ARXIV:0801.0450;%%

\bibitem{KRprep}
  I.~Kuznetsova and J.~Rafelski,
in preparation.

%\cite{Torrieri:2006yb}
\bibitem{Torrieri:2006yb}
  G.~Torrieri and J.~Rafelski,
  %``Hadron resonances and phase threshold in heavy ion collisions,''
  Phys.\ Rev.\  C {\bf 75}, 024902 (2007)
  [arXiv:nucl-th/0608061].
  %%CITATION = PHRVA,C75,024902;%%

 \end{thebibliography}
\end{document}